\documentclass[aps,amsfonts,amsmath,prd,nofootinbib,twocolumn]
{revtex4-1}
\usepackage[latin1]{inputenc}
\usepackage{amssymb}
\usepackage{amsmath}
\usepackage{amsfonts}
\usepackage{enumerate}
\usepackage{epsfig}
\usepackage[pdftex]{hyperref}
\usepackage{subfigure,slashed}
\usepackage{comment}
\usepackage{xcolor}

\newcommand{\red}[1]{{\textcolor{black} {#1}}}
\newcommand{\RED}[1]{{\textcolor{black} {#1}}}
\newcommand{\be}{\begin{eqnarray}}
\newcommand{\ee}{\end{eqnarray}}
\def\ba{\begin{array}}
	\def\ea{\end{array}}

\makeatletter

\def\ba{\begin{eqnarray}}
\def\ea{\end{eqnarray}}
\def\beas{\begin{eqnarray*}}
	\def\eeas{\end{eqnarray*}}
\def\nn{\nonumber}
\def\d{\partial}
\def\sla{\raise.15ex\hbox{$/$}\kern-.57em}

\newcommand\eps{\epsilon}

\begin{document}

	\title{\Large Hidden and explicit quantum scale invariance}
	
	\author{Sander Mooij}
\email{sander.mooij@epfl.ch}

\author{ Mikhail Shaposhnikov}

\email{mikhail.shaposhnikov@epfl.ch}

\author{Thibault Voumard}
\email{thibault.voumard@alumni.epfl.ch}

	\affiliation{Institute of Physics, Laboratory for Particle Physics and Cosmology (LPPC),\\
		\'Ecole Polytechnique F\'ed\'erale de Lausanne (EPFL), CH-1015 Lausanne, Switzerland}

	\begin{abstract}
		There exist renormalisation schemes that explicitly preserve the scale invariance of a theory at the quantum level. Imposing a scale invariant renormalisation breaks renormalisability and  induces new non-trivial operators in the theory. In this work, we study the effects of such scale invariant renormalisation procedures. On the one hand, an explicitly quantum scale invariant theory can emerge from the scale invariant renormalisation of a scale invariant Lagrangian. On the other hand, we show how a quantum scale invariant theory can equally emerge from a Lagrangian visibly breaking scale invariance renormalised with scale dependent renormalisation (such as the traditional $\overline{\rm MS}$ scheme). In this last case, scale invariance is  hidden in the theory, in the sense that it only appears explicitly after renormalisation.
	\end{abstract}

	\maketitle

	\section{Introduction}
	
	It is a textbook truth that a classical theory is defined completely by its Lagrangian whereas for the formulation of a quantum field theory (QFT) one needs a Lagrangian and a re\-nor\-ma\-li\-sa\-tion procedure for re\-mo\-ving the divergences. This combination can render the symmetries of the QFT obscure: if a Lagrangian possesses a symmetry but the re\-nor\-ma\-li\-sa\-tion scheme does not preserve this symmetry, the renormalised theory is not explicitly symmetric. This is the case with scale invariance (SI)\footnote{The SI theories may be interesting from the point of view of the hierarchy problem, see e.g.  \cite{Wetterich:1983bi, Shaposhnikov:2007nj,Shaposhnikov:2008xi,Shaposhnikov:2008xb,Bardeen:1995kv,Wetterich:2011aa,Foot:2007iy,Giudice:2008bi,Vissani:1997ys,Farina:2013mla,Heikinheimo:2013fta,Carone:2013wla,Quiros:2013mza,Meissner:2006zh,AlexanderNunneley:2010nw,Hur:2011sv,Wetterich:2014gaa,Ferreira:2016vsc,Ferreira:2016wem,Grabowski:2018fjj, Karam1, Karam2, Anupam}.}, where, for example,  the $\overline{\rm MS}$ subtraction scheme \cite{tHooft:1972tcz} breaks SI explicitly by introducing an explicit mass scale $\mu$. One may circumvent this problem by using renormalisation procedures designed to preserve the symmetries of a given Lagrangian. Renormalisation schemes that preserve SI have been proposed and give rise to the cancellation of the dilatation anomaly \cite{Englert:1976ep,Wetterich:1987fm} \cite{Shaposhnikov:2008xi,Shaposhnikov:2008ar}. \red{In particular, in \cite{Roberta} an approach was presented to preserve conformal symmetry up to first order in perturbation theory. This result has been extended to any fixed loop order in the elegant work \cite{Sacha}. It was shown that one can choose conformally invariant counterterms such that the re\-sul\-ting quantum corrected theory is conformally invariant and finite. } The simplest formulation \red{of scale invariant renormalisation} uses dimensional regularisation, but with the renormalisation scale $\mu$ replaced by a field o\-pe\-ra\-tor of the appropriate mass dimension. No explicit mass scale is introduced in the theory, and SI is preserved.
	
	These SI schemes thus have the peculiarity of using dynamical fields to generate renormalisation scales. This leads to a conflict between renormalisability and scale invariance and thus to the introduction of infinite series of operators into the Lagrangian which ultimately ensure scale invariance of the quantum theory.  The aim of this work is to show how one can trade these dynamical scales for some specific modification of the Lagrangian con\-si\-der\-ed, while equipping it with the traditional, constant renormalisation scale $\mu$.
	
	This is made possible by the following obvious observation. If one starts from a given SI Lagrangian equipped with a dynamical renormalisation scale, one can derive a set of Green's functions at any chosen loop level. This set of Green's functions defines the theory (at the chosen loop level). However, the Lagrangian of the QFT that produces this set of Green's functions is not unique, as a change in the renormalisation procedure can be compensated by an appropriate modification of the Lagrangian. In particular, starting from a SI Lagrangian equipped with SI preserving renormalisation, one can always find a new Lagrangian equipped with traditional scale dependent renormalisation that will produce the same set of Green's functions at the chosen loop level. Clearly, the Lagrangian in this new theory will need to break SI, in order to compensate for the SI breaking coming from the scale dependent renormalisation. The original symmetry of the theory is then hidden in the new Lagrangian, but becomes explicit again when the theory is renormalised at the chosen loop level with the symmetry breaking scheme. We will explicitly construct this new Lagrangian for three examples, at the one-loop level. 


	
	This paper is organised as follows. We begin by reminding the reader about scale invariant renormalisation in Section \ref{sir}. The first example, given in Section \ref{doublescalar}, uses the simplest non-trivial scenario one could have, i.e. a theory with two scalar fields. The second one includes a fermion field in order to show how the concept ge\-ne\-ra\-li\-ses to theories with wave-function renormalisation and is discussed in Section \ref{yukawa}. We finally discuss the inclusion of gravity in a model with non-minimal coupling between gravity and the scalar sector of the SM in Section \ref{higgsinflation}. We present our conclusions in Section \ref{summary}.
	
	\section{\label{sir}Scale invariant renormalisation}
	
	Suppose taking a classically SI theory, that is, an action invariant under a rescaling of the coordinates $x^{\mu}\rightarrow \alpha x^{\mu}$, accompanied by a rescaling of the fields $\phi \rightarrow \alpha^{-\Delta_{\phi}} \phi$, where $\Delta_{\phi}$ is the canonical mass dimension of the field $\phi$. It should be noted that this symmetry is satisfied as long as no explicit massive {(dimensionful)} parameter enters the theory, as these do not get rescaled by the dilatation. {However}, when renormalising the theory using, for example, $\overline{\rm MS}$ dimensional regularisation, one {precisely} introduces such a mass scale $\mu$ to regularise the divergences. This mass {scale} explicitly breaks SI.

	To remedy this situation \cite{Englert:1976ep},\cite{Shaposhnikov:2008xi}, one replaces $\mu$ by an operator with the appropriate mass dimension, as fields will rescale under a dilatation. No explicit mass pa\-ra\-me\-ter then enters the QFT, thus SI is protected at the quantum level.
In our case, for the sake of simplicity, we will only use scalar fields. Supposing that we have a theory with a scalar field $\sigma$ (hereafter, the dilaton), we can choose in $d$-dimensions
	\be 
	\label{musigma} \mu \propto \sigma^{2/(d-2)}.
	\ee
	In order to use this scale for perturbative computations, $\sigma$ is required to have a VEV $\bar{\sigma}$. If it does not, eq.~\ref{musigma} admits no polynomial expansion \RED{and the dilaton field cannot be used for perturbative renormalisation}. Using a more physical point of view, the VEV is required to ge\-ne\-rate the massive scale necessary for renormalisation and to reproduce the original running of the couplings \cite{Shaposhnikov:2008xi,Tamarit:2013vda}. \RED{Without the dilaton taking a vev, perturbative scale invariant renormalisation is impossible. It might be that in the absence of a dilaton vev scale invariant renormalisation can be done non-perturbatively, but this is not at all clear yet. See for example \cite{tkachev} for a lattice approach. On the other hand, it has been shown in \cite{Shaposhnikov:2008xb,GarciaBellido:2011de,Ferreira:2016vsc} that once gravity is included, there is always a solution with $\bar{\sigma}\neq 0$.}
	
\RED{We note that the dilaton field is an extra field that is added to the model in order to facilitate scale invariant renormalisation. This does not limit the type of interactions that can be renormalised with the dilaton. Any interaction that in dimensional regularisation gets renormalised with the help of $\mu$, now gets renormalised with the help of the dilaton's vev $\bar{\sigma}$.}

	Although this scale ensures a SI form of the quantum corrections, it has other non-trivial effects. As in usual dimensional regularisation, we must redefine the couplings of the theory to keep them dimensionless in $d=4-2\epsilon$ dimensions using the renormalisation scale. This induces new couplings between the dilaton and the other fields. For example, taking the simple theory
	\be 
	S=\int d^4x \left[\frac{1}{2}\partial_{\mu}h\partial^{\mu}h -\frac{\lambda h^4}{4!} + \frac{1}{2}\partial_{\mu}\sigma\partial^{\mu}\sigma \right], 
	\ee
	one makes the replacement in $d=4-2\epsilon$ dimensions, using $\sigma=\bar{\sigma}+\hat{\sigma}$
	\ba
	\label{lambdaredef} \lambda \rightarrow \sigma^{2\epsilon/(1-\epsilon)} \lambda &=& \lambda \bar{\sigma}^{2\epsilon}\left[1+2\epsilon \sum_{n=1}^{\infty} \frac{(-1)^{n-1}}{n}\left(\frac{\hat{\sigma}}{\bar{\sigma}}\right)^n\right] \nn\\
&& \qquad + \mathcal{O}(\epsilon^2). 
	\ea
	One sees a major consequence of taking a dynamical renormalisation scale: it induces infinite series of new operators in the Lagrangian, suppressed by the VEV of the dilaton.

 \RED{Note that for this series to make sense, the fluctuations $\hat{\sigma}$ need to be small compared to the background field value $\bar{\sigma}$. This is reflected in the earlier observation that scale independent renormalisation breaks renormalizibility. It generates operators of mass dimension higher than four, suppressed by appropriate powers of the background field. The demand that these higher order operators do not spoil the predictivity of the theory is precisely equivalent to the demand that the series expansion in eq.~\ref{lambdaredef} is sensible.}
	
	Consequently, equipping a Lagrangian with a dy\-na\-mi\-cal renormalisation scale can preserve SI but renders the theory non-renormalisable \cite{Englert:1976ep}. For a potentially renormalisable Lagrangian, choosing the scale invariant prescription then amounts to trading re\-nor\-ma\-li\-sa\-bi\-li\-ty for scale invariance at the quantum level, provided the original Lagrangian is SI. On the other hand, if the Lagrangian is not scale invariant at the classical level, then it seems that the scale invariant prescription would never be of use, as no symmetry is preserved and re\-nor\-ma\-li\-sa\-bi\-li\-ty is lost. However, this prescription takes all its meaning when gravity is taken into account. In this case, the theory is always non-renormalisable, so no desirable property of the theory is lost by choosing a scale invariant prescription. 
	
	The infinite series of operators appearing during the regularisation make it clear that, beginning from a given Lagrangian, choosing a dynamical renormalisation scale over a traditional, constant scale amounts to con\-si\-de\-ring a different QFT. As explained before, we will show that one can always translate this dynamical scale into a mo\-di\-fi\-ca\-tion of the Lagrangian. The new Lagrangian is constructed in such a way that, when renormalised at a constant scale, it gives rise to the same Green's functions produced by the original QFT at a chosen loop level. In the next Sections, we explicitly show how this can be done at the one-loop level for a scalar theory, a Yukawa theory and a theory where gravity is included, and discuss the generalisation to any loop level.

	\section{\label{doublescalar}Scalar field theory}
	
	Let us consider the simple model with two scalar fields described by the scale invariant Lagrangian
	\be 
	\mathcal{L} = \frac{1}{2}\partial_{\mu}h\partial^{\mu}h + \frac{1}{2}\partial_{\mu}\sigma\partial^{\mu}\sigma -\frac{\lambda}{4!}h^4. 
	\ee
	To get a SI quantum theory, we choose a renormalisation scale $\mu = \sigma^{2/(d-2)}$, where $d$ is the spacetime dimension. As stated before, in order to make a perturbative expansion, we must impose that SI is broken spontaneously so that the field $\sigma$ has a non-zero VEV $\bar{\sigma}$, i.e. the potential must have a flat direction.
	
	As we expect no wave-function renormalisation at one-loop for scalar fields, it is sufficient here to consider the effective potential \cite{Coleman:1973jx,Jackiw:1974cv}. To continue the theory in $d=4-2\epsilon$ dimensions, we replace
	\be 
	\lambda \rightarrow \mu^{2\epsilon} \lambda, 
	\ee
	which ensures that the coupling $\lambda$ remains dimensionless. {When expanding $\mu^{2\eps} =\sigma^{2\eps/(d-2)}$, we now see that} this redefinition introduces an infinite number of new operators in the theory. Still, as can be seen from eq.~\ref{lambdaredef}, only the lowest order operator in $\hat{\sigma}/\bar{\sigma}$
	\be
	\label{lowestop} \lambda \bar{\sigma}^{2\epsilon} \frac{h^4}{4!}
	\ee
	comes at order $\mathcal{O}(\epsilon^0)$. All the operators coming at $\mathcal{O}(\epsilon)$ order are called evanescent, as they disappear when setting $d=4$. The one-loop effective potential is then
	\ba
	{V_1}&=&\mu^{2\epsilon}\left[\frac{\lambda h^4}{4!}+\frac{\lambda^2h^4}{(16\pi)^2}\left(-\frac{1}{\epsilon}+\ln\left(\frac{\lambda h^2}{8\pi e^{3/2-\gamma}\mu^2}\right)\right)\right] \nn\\
&& \qquad +\mathcal{O}(\epsilon), 
	\ea
	where the $\mathcal{O}(\epsilon)$ term accounts for the contribution of the whole evanescent infinite series. When we send $\epsilon \rightarrow 0$, the contribution of the infinite series thus vanishes. It is important to note that this is only true at the one-loop level. At higher loop levels, divergences of order $\mathcal{O}(\epsilon^{-2})$ can appear, and the contribution of $\mathcal{O}(\epsilon)$ terms in the effective potential does no longer vanish when $\epsilon \rightarrow 0$ \cite{Tamarit:2013vda, Ghilencea:2017yqv}. \RED{However, let us stress again that, notwithstanding the increasing complexity, scale invariant renormalisation is well defined up to any fixed loop order, as shown in \cite{Sacha}. Explicit results up to third order in the loop expansion can be found in \cite{Ghilencea:2017yqv}.}

	We regularise the divergence by adding a counter-term
	\be 
	\label{deltalambdadouble} \mu^{2\epsilon}\delta_{\lambda} h^4,\ \ \ \ \ \delta_{\lambda}=\frac{\lambda^2}{(16\pi)^2}\left(\frac{1}{\epsilon}-\ln\left(\frac{\lambda}{8\pi e^{3/2-\gamma}}\right)\right). 
	\ee
	This particular choice of counter-term recasts the potential into a simple form when $\epsilon \rightarrow 0$
	\be 
	\label{effpotdouble} V_1= \frac{\lambda h^4}{4!} + \frac{\lambda^2 h^4}{(16\pi)^2}\ln\left(\frac{h^2}{\sigma^2}\right). 
	\ee
	As one can see, the one-loop effective potential\footnote{
		{We will sloppily refer to the potential in eq.~\ref{effpotdouble} as ``effective potential", even if it contains not only background fields, but also quantum fluctuations.}
	} is 
	ex\-pli\-cit\-ly SI{: it is free of explicit mass scales. This would not be so had we used traditional dimensional regularisation.} A result for more general potentials following the same approach has been given in \cite{Ghilencea:2015mza}. 
	
	In order to clarify the effects of the dynamical nature of the renormalisation scale, we now show that we can start from a different potential equipped with another renormalisation scale, i.e. another QFT, and end up with the same effective potential. Here we choose the simplest renormalisation scale one could imagine, a constant one. {In other words: we return to traditional dimensional re\-gu\-la\-ri\-sa\-tion.} We must then construct a new Lagrangian which will give rise to eq.~\ref{effpotdouble} when renormalised at this constant scale at the one-loop level.
	
	We required at the beginning that the SI must be spontaneously broken, so let us expand eq.~\ref{effpotdouble} around the VEV of $\sigma = \bar{\sigma}+\hat{\sigma}$
	\ba
	\label{effpotsc} V_1&=& \frac{\lambda h^4}{4!} + \frac{\lambda^2 h^4}{(16\pi)^2}\ln\left(\frac{h^2}{\bar{\sigma}^2}\right)-\frac{2\lambda^2 h^4}{(16\pi)^2}\ln\left(1+\frac{\hat{\sigma}}{\bar{\sigma}}\right). \nn\\
	\ea
	One can identify two distinct contributions in eq.~\ref{effpotsc}: the first two terms correspond to a $\lambda h^4$ theory renormalised at a constant scale $\mu=\bar{\sigma}$, whereas the last term corresponds to an infinite series of operators which ensures SI of the total potential. From this observation, one can guess that the potential (where the second term carries a factor $\hbar$ that we are suppressing in this work)
	\be 
	\label{fundpotss} \tilde{V_0} = \frac{\lambda h^4}{4!} -\frac{2\lambda^2 h^4}{(16\pi)^2}\ln\left(1+\frac{\hat{\sigma}}{\bar{\sigma}}\right) 
	\ee
	will give rise to eq.~\ref{effpotdouble} when renormalised at the constant scale $\bar{\sigma}$.  Indeed, using this constant scale and this new potential, one gets for the one-loop effective potential
	\ba
	\tilde{V_1}&=&\mu^{2\epsilon}\Biggl[\frac{\lambda h^4}{4!}-\frac{2\lambda^2 h^4}{(16\pi)^2}\ln\left(1+\frac{\hat{\sigma}}{\bar{\sigma}}\right)\nn\\
&& \qquad+\frac{\lambda^2 h^4}{(16\pi^2)}\left(\frac{-1}{\epsilon}+\ln\left(\frac{\lambda h^2}{8\pi e^{3/2-\gamma}\bar{\sigma}^2}\right)\right)\Biggr] \nn\\
&& \qquad+ \mathcal{O}(\lambda^3) + \mathcal{O}(\epsilon).
	\ea
	By sending $\epsilon \rightarrow 0$ and acknowledging that our expansion parameter is $\lambda$, i.e. that one-loop corrections are relevant up to $\mathcal{O}(\lambda^2)$, one gets the same effective potential as in eq.~\ref{effpotdouble} by choosing the same $\delta_{\lambda}$ counter-term as in eq.~\ref{deltalambdadouble}.
	
	Using this construction, we see that taking a dynamical renormalisation scale amounts to choosing a Lagrangian containing an infinite series of operators suppressed by the VEV of the dynamical renormalisation scale. Consequently, when the VEV of the dynamical renormalisation scale is large, taking the dynamical or the constant scale yields the same physical predictions. On the other hand, the physics differ when considering large field fluctuations.
	
	It is interesting to note that $V_0'$ is not SI, but when renormalised at the scale $\bar{\sigma}$, one recovers a SI quantum corrected theory. The SI of the theory is thus hidden at the tree level, and made explicit after (one-loop) renormalisation: the scale dependence of the quantum corrections compensates the scale dependence of the tree level Lagrangian. Note however that to find the hidden scale invariance in eq.~\ref{fundpotss}, one needs to take every term into account; an EFT approach based on {a truncation at a certain order in $1/\bar{\sigma}$ does not suffice.} Only the total expression in eq.~\ref{fundpotss}, as opposed to every individual order in $1/\bar{\sigma}$, is scale invariant.
	
	The generalisation of the construction to other potentials is straightforward, the procedure being similar. It is clear that this construction can be applied at any loop level, although it becomes more complicated as early as at the two-loop level. As mentioned before, divergences of order $\mathcal{O}(\epsilon^{-2})$ lead to non-vanishing contributions com\-ing from the evanescent operators of eq.~\ref{lambdaredef} \cite{Tamarit:2013vda, Ghilencea:2017yqv}. Then, the same contributions should be added to the new Lagrangian to reproduce the quantum result, in a way si\-mi\-lar to what we did at the one-loop level. One also needs to compensate the corrections generated by the terms needed to reproduce the lower loop levels results, which can always be done by adding appropriate operators of higher order in the couplings \RED{\cite{Sacha}}.
	
	\section{\label{yukawa}Yukawa theory}
	
	To generalise the example we proposed in the last Section, we now consider a model which {requires} a non trivial wave-function renormalisation at {the} one-loop level. The Lagrangian we write contains a fermion and a dilaton:
	\be 
	\mathcal{L} = \frac{1}{2}\partial_{\mu}\sigma\partial^{\mu}\sigma + i\bar{\Psi}\slashed{\partial}\Psi -\frac{\lambda \sigma^4}{4!}-y\bar{\Psi}\sigma\Psi. 
	\ee
	This is the most general SI Lagrangian containing a scalar field and a fermion {with field operators of mass dimension less or equal to 4}.
	
	As before, we require the SI to be spontaneously broken, although we do not specify how this effect occurs in this particular model. {We simply assume that} the scalar field $\sigma$ 
	acquires a VEV $\bar{\sigma}$, and we define a dynamical renormalisation scale as in the previous model,
	\be 
	\mu = \sigma^{\frac{2}{d-2}}.
	\ee
	This mass scale is used to define dimensionless couplings in dimension $d=4-2\epsilon$ :
	\be 
	\lambda \rightarrow \mu^{2\epsilon}\lambda ~ \RED{=~\sigma^{\frac{2\eps}{1-\eps}}\lambda}\ee
	\be y \rightarrow \mu^{\epsilon}y\RED{~=~\sigma^{\frac{\eps}{1-\eps}}y}. 
	\ee
	The $d=4-2\epsilon$ Lagrangian then reads
	\be 
	\mathcal{L} = \frac{1}{2}\partial_{\mu}\sigma\partial^{\mu}\sigma + i\bar{\Psi}\slashed{\partial}\Psi -\frac{\lambda \sigma^{4+2\epsilon}}{4!}-y\bar{\Psi}\sigma^{1+\epsilon}\Psi, 
	\ee
	where we have dropped $\mathcal{O}(\epsilon^2)$ terms as they will not contribute to the final results at one-loop.
	
	To obtain suitable Feynman rules, one has to expand $\sigma$ around its VEV: $\sigma = \bar{\sigma} + \hat{\sigma}$. One can check that this generates a finite number of simple interactions of order $\mathcal{O}(\epsilon^0)$, and an infinite number of interactions of order $\mathcal{O}(\epsilon)$ and higher. By recollecting every superficially divergent diagram, one can see that, at the one-loop level, up to the freedom in the definition of the counter-terms, i.e. up to finite contributions, the theory that we need to renormalise is equivalent to
	\be 
	\label{Lpsisig} \mathcal{L}= \frac{1}{2}\partial_{\mu}\sigma\partial^{\mu}\sigma + i\bar{\Psi}\slashed{\partial}\Psi -\frac{\lambda\bar{\sigma}^{2\epsilon} \sigma^4}{4!}-y\bar{\sigma}^{\epsilon}\bar{\Psi}\sigma\Psi, 
	\ee
	where the $\epsilon$ powers of the dynamical field are no longer present. Consequently, at the one-loop level, considering the renormalisation scale constant during regularisation only induces finite errors that can be absorbed in the counter-terms. One should note that this simplification can not be done at higher loop levels, as divergences of order $\mathcal{O}(\epsilon^{-2})$ or lower are generated, {which} multiply the interactions of order $\mathcal{O}(\epsilon)$.
	
	This theory is well known, and renormalisation is now straightforward. In the same way that an effective potential can be defined, we define an effective Lagrangian from the one given in eq.~\ref{Lpsisig}. This effective Lagrangian has the properties that its diagrams at tree level give the same results as those of the starting Lagrangian at one-loop. The effective Lagrangian which satisfies this requirement is (as $\epsilon \rightarrow 0$) in momentum space
	\ba \label{L1}
	\mathcal{L}_1 &=& \mathcal{L} +\frac{y^2}{3(4\pi)^2}p^2\sigma^2 \ln\left(\frac{-p^2}{\sigma^2}\right) - \frac{y^2}{2(4\pi)^2}\bar{\Psi}\slashed{p}\Psi\ln\left(\frac{-p^2}{\sigma^2}\right) \nn\\
&& \qquad -\frac{\sigma^4}{(16\pi)^2}(\lambda^2-16y^4)\ln\left(\frac{-p^2}{\sigma^2}\right)\nn\\
&& \qquad-\frac{y^3}{(16\pi)^2}\bar{\Psi}\sigma\Psi \ln\left(\frac{-p^2}{\sigma^2}\right).
	\ea
	This result is SI, which means that choosing a dynamical regularisation scale to preserve SI not only works for effective potentials, but also works for full Lagrangians, as should be expected from the absence of an explicit mass scale.
	
	We now repeat the construction done in the previous Section, that is we construct a new QFT which leads to the same effective Lagrangian, but with a constant renormalisation scale. This allows us to translate the dynamical nature of the renormalisation scale into a choice of tree level Lagrangian, while leaving the physical predictions untouched.
	
	To do so, one expands eq.~\ref{L1} around the VEV of $\sigma$. One piece will correspond to the effective Lagrangian obtained by renormalising eq.~\ref{Lpsisig} at a constant scale $\mu=\bar{\sigma}$, the other piece comes from the expansion around the VEV. Consequently, in all similarity with the previous Section, we propose for the new Lagrangian
	\ba  \label{LC}
	\tilde{\mathcal{L}}&=& \mathcal{L} -\partial_{\mu}\hat{\sigma}\partial^{\mu}\hat{\sigma} \frac{2y^2}{3(4\pi)^2}\ln\left(1+\frac{\hat{\sigma}}{\bar{\sigma}}\right)\nn\\
&&\qquad + \bar{\Psi}\slashed{\partial} \Psi \frac{iy^2}{(4\pi)^2}\ln\left(1+\frac{\hat{\sigma}}{\bar{\sigma}}\right)\nn\\
&& \qquad+\frac{2(\hat{\sigma}+\bar{\sigma})^4}{(16\pi)^2}(\lambda^2-16y^4)\ln\left(1+\frac{\hat{\sigma}}{\bar{\sigma}}\right) \nn\\
&& \qquad+ \frac{2y^3}{(16\pi)^2}\bar{\Psi}\Psi(\hat{\sigma}+\bar{\sigma})\ln\left(1+\frac{\hat{\sigma}}{\bar{\sigma}}\right),
	\ea
	which is the original Lagrangian to which the contribution coming from the dynamical scale has been explicitly added. This Lagrangian, by the same kind of arguments that were given in the example with the scalar fields, will give rise to eq.~\ref{L1} when renormalised at one-loop, at the scale $\bar{\sigma}$. Corrections coming from the series term are of too high order in the couplings to be relevant at one-loop level, and can thus be dropped. Once again, we see that choosing a field-dependent renormalisation scale is e\-qui\-va\-lent to adding infinite series of operators suppressed by the VEV of the dilaton. Moreover, we also see that SI is hidden in the new Lagrangian, but is recovered at the one-loop level.

\RED{We note that if we would include an extra scalar field $h$ to this model, with an additional Yukawa interaction $y_h \bar{\Psi} h \Psi$, SI renormalisation works out in exactly the same way. Again we set $y_h \to \sigma^{\frac{\eps}{1-\eps}}y_h$ to define a dimensionless coupling in $d=4-2\eps$. It is by no means necessary that the interaction-to-be-renormalised already contains the dilaton field in $d=4$.}

This construction can be done at any loop level, and for any theory admitting a perturbative expansion, although it becomes complicated at the two-loop level already, for the same reasons that were discussed in Section \ref{doublescalar}. 

\RED{We note that if we consider a similar Yukawa interaction, but now including $\gamma_5$, we run in the same problem of how to define $\gamma_5$ in $d=4-2\eps$ as in dimensional regularisation \cite{tHooft:1972tcz}.}

\RED{Let us finish this section with a quick look at scale invariant renormalisation when gauge fields are present:
\be
\mathcal{L} \supset -\frac{1}{4} F_{\mu\nu} F^{\mu\nu} + \bar{\Psi} \gamma^\mu\left(\d_\mu -ig A_\mu\right)\phi \Psi.
\ee
Clearly, multiplying $g$ by an appropriate power of the dilaton breaks gauge invariance. The solution is to rescale the gauge field such that we get
\be
\mathcal{L} \supset -\frac{1}{4g^2} F_{\mu\nu} F^{\mu\nu} + \bar{\Psi} \gamma^\mu\left(\d_\mu -i A_\mu\right)\phi \Psi.
\ee
Now the gauge field has mass dimension $1$, in any number of dimensions. In this last theory, we can regularize in $d=4-2\eps$ by setting $g^2 \to \sigma^{\frac{2\eps}{1-\eps}}g^2$. With this explicitly gauge invariant and scale invariant regularisation, one is guaranteed that, just like after dimensional regularisation, the resulting effective potential is gauge invariant on-shell, up to any order in the loop expansion.}


	\section{\label{higgsinflation}Inclusion of gravity}

	In this Section, we present an explicit example of our construction in a model with gravity. As already mentioned in the introduction, theories including gravity are non-renormalizable from the beginning. Therefore, we lose nothing in applying scale invariant renormalisation. Let us consider a scale invariant modification of the SM coupled in a non-minimal way to gravity \cite{Englert:1976ep},\cite{Shaposhnikov:2008xi}:
	\ba	\mathcal{L} &=& -\left(\xi_\sigma \sigma^2 + \xi h^2\right)\frac{R}{2}
	+\frac{1}{2}\left[\left(\d_\mu\sigma\right)^2+\left(\d_\mu h\right)^2\right]\nn\\
&& \qquad
	-\lambda\left(h^2-\zeta^2 \sigma^2\right)^2 +\mathcal{L}_{SM}.\ea
	Here, $\sigma$ denotes the dilaton field, $h$ is the Higgs field, $\xi_\sigma$, $\xi$, $\lambda$ and $\zeta$ are dimensionless coupling constants, and $\mathcal{L}_{\rm SM}$ contains all Standard Model terms apart from the pure scalar sector. For phenomenology, the Higgs-dilaton potential can be chosen such that after spontaneous symmetry breaking it reduces to the usual SM potential, while the first term in this Lagrangian effectively plays the role of the Planck constant: $\xi_\sigma \bar{\sigma}^2 \to M_p^2$. For simplicity, we neglect further effects of the dilaton. We are thus left after spontaneous symmetry breaking with a theory whose gravity and scalar sectors read
	
	\ba
	\label{actionjordan} S_J&=&\int d^4x \sqrt{-g}\Biggl[-\frac{M_p^2+\xi h^2}{2}R + \frac{1}{2}\partial_{\mu}h\partial^{\mu}h\nn\\
&&\qquad \qquad\qquad-\frac{\lambda}{4}(h^2-v^2)^2\Biggr].
	\ea
	Note that this Lagrangian can describe Higgs inflation (when $\xi \gg 1$) \cite{Bezrukov:2007ep,Salopek:1988qh}.
	
	One can get rid of the non-minimal coupling of the Higgs field to gravity by making a conformal transformation to the Einstein frame
	\be 
	\label{metricrescale} \hat{g}_{\mu\nu}=\Omega^2g_{\mu\nu}, 
	\ee
	with
	\be 
	\Omega^2=\frac{M_p^2+\xi h^2}{M_p^2}. 
	\ee
	This transformation yields a non-canonical kinetic term for the Higgs field. A scalar field $\chi$ with canonical kinetic term can be defined as
	\be 
	\label{higgschange}  \frac{d\chi}{dh}=\sqrt{\frac{\Omega^2+6\xi^2h^2/M_p^2}{\Omega^4}}, 
	\ee
	giving an Einstein frame action
	\be 
	S_E=\int d^4x \sqrt{-\hat{g}}\left[-\frac{M_p^2}{2}\hat{R}+\frac{1}{2}\partial_{\mu}\chi\partial^{\mu}\chi-V_0(\chi)\right]
	\ee
	where $\hat{R}$ is computed using $\hat{g}_{\mu\nu}$ and
	\be 
	V_0(\chi) = \frac{\lambda}{4\Omega(\chi)^4}\left(h(\chi)^2-v^2\right)^2. 
	\ee
	
	Notice that the conformal transformation eq.~\ref{metricrescale} does not only change the gravitational part of the Lagrangian, but also induces changes in the other terms. The net effect is a $\Omega^{-1}$ rescaling of every mass scale in the theory. Supposing that we renormalise the Jordan frame Lagrangian {at} a scale $\mu_J$, the equivalent scale in the Einstein frame is then $\mu_E=\mu_J/\Omega(\chi)$, making the re\-nor\-ma\-li\-sa\-tion scale field dependent. A natural choice for the re\-nor\-ma\-li\-sa\-tion scale appears in our theory: the Planck mass $M_p$. We thus get two different prescriptions for renormalisation, depending which frame is chosen to have the natural scale. The first one takes a constant scale in the Einstein frame,
	\be 
	\text{Prescription I :} \label{prescription1}\ \  \ \mu_J^2 = M_p^2+\xi h^2, \  \ \mu_E^2=M_p^2,
	\ee
	whereas the second one uses the constant scale in the Jordan frame
	\be 
	\text{Prescription II :}\ \  \ \mu_J^2 = M_p^2, \  \ \mu_E^2 = \frac{M_p^4}{M_p^2+\xi h^2}. 
	\ee
	It is not clear which choice should be used without knowing physics at the Planck scale \cite{Kengo}. As we see, one always produces a dynamical renormalisation scale when renormalising, be it in the Jordan or in the Einstein frame. Still, as long as one sticks to one definite prescription, physics will always be the same in the Jordan and in the Einstein frame \cite{George:2013iia,Postma:2014vaa,Makino:1991sg,Karamitsos:2017elm}\footnote{\red{Note however that even if functional prescriptions (as a function of $N$) for all inflationary observables are equal in both frames, evaluating them at a given number of e-folds before the end of inflation does not have the same meaning in both frames: $N_J \neq N_E$ \cite{Karam3}.}}. We can thus choose the frame in which we want to work. In our case, {we choose for} the Einstein frame, as gravity is in canonical form and its back-reaction can be neglected in front of the corrections coming from the SM loops \cite{George:2015nza,Bezrukov:2009db}. We will repeat here the constructions done in the previous Sections to clarify the differences between the two prescriptions, by translating the field dependent renormalisation scale into a choice of the potential. This allows one to directly see the differences that arise from choosing one or the other prescription.

	The effective potential for the Einstein frame has been computed at one-loop in \cite{Bezrukov:2008ej}: counting the Higgs, two W's, the Z and the three colours of the top quark, one obtains, {up to finite terms,}
	\ba
	\label{einsteinonelooppot} V_1 &=& V_0 + \frac{m_H^4}{64 \pi^2}\ln\left(\frac{m_H^2}{\mu^2}\right)+ \frac{6 m^4_W}{64\pi^2}\ln\left(\frac{m^2_W}{\mu^2}\right) \nn\\
&& \qquad+ \frac{3 m^4_Z}{64\pi^2}\ln\left(\frac{m^2_Z}{\mu^2}\right) -\frac{3 m_t^4}{16\pi^2}\ln\left(\frac{m_t^2}{\mu^2}\right), 
	\ea
	where $m$ stands for the mass of the subscripted particle {field} in the Einstein frame and $\mu^2=M_p^4/(M_p^2+\xi h^2)$. Notice that the contribution coming from the other quarks is negligible in front of $m_t$. It is worth noting that although $\mu^2$ is not purely dynamical
, the regularisation procedure is still well defined. Notice that here we took the liberty of considering $\mu$ as field independent when computing the effective potential. {This amounts to neglecting finite corrections in the final result, coming from the infinite series of operators of order $\mathcal{O}(\epsilon)$ which appears when expanding the dynamical scale around $M_p$.} As {we have seen in the previous Sections, this can be done only at the one-loop level. At higher loop levels, divergences of higher order lead to infinite contributions coming from the evanescent operators.}
	
	Noticing that
	\ba
	\ln\left(\frac{m^2}{M_p^4/(M_p^2+\xi h^2)}\right) &=& \ln\left(\frac{m^2}{M_p^2}\right) \nn\\
&& + \sum_{n=1}^{\infty} \frac{(-1)^{n-1}}{n}\left(\frac{\xi h^2}{M_p^2}\right)^n 
	\ea
	and using the results obtained in the preceding Sections, one can deduce that a physically equivalent potential is
	\ba
	\label{poteinstein} \tilde{V_0} &=& V_0 + \frac{1}{64\pi^2}(m_H^4+ 6 m_W^4 + 3 m_Z^4 -12 m_t^4)\nn\\
&& \qquad\qquad \times\sum_{n=1}^{\infty} \frac{(-1)^{n-1}}{n}\left(\frac{\xi h^2}{M_p^2}\right)^n, 
	\ea
	equipped with the constant renormalisation scale $\mu^2=M_p^2$.
	This potential, when renormalised at the scale $\mu^2=M_p^2$ in the Einstein frame gives rise to eq.~\ref{einsteinonelooppot}.
	
	From eq.~\ref{poteinstein}, one sees that choosing one or the other prescription amounts to adding an infinite series of o\-pe\-ra\-tors suppressed by powers of $M_p/\sqrt{\xi}$ in the Lagrangian. {If we choose $\xi \gg 1$, as in Higgs inflation, we see that,} consequently, both prescriptions correspond to identical Lagrangians in the small field regime {($h < M_p/\xi$)}, but not in the large field regime {($h>M_p/\sqrt{\xi}$)}, leading to the different predictions related to the choice of prescription. Even though both frames are physically equivalent, this construction shows that taking {the renormalization scale} constant in one frame, {or in another frame,} leads to different physics. This result is comparable to what has been found in \cite{Fumagalli:2016lls} and \cite{Hamada:2016onh}. See also \cite{Mario}\red{, which discusses the choice for a prescription in terms of the path integral measure: a constant renormalisation scale corresponds to a trivial integration measure}. {We emphasize that this \red{``prescription dependence"} is not a manifestation of the breaking of frame invariance: as we have just shown, taking the re\-nor\-ma\-li\-za\-tion scale constant in frame A, or taking it constant in frame B, describes two physically different situations that can both equivalently be studied in whatever frame one prefers.}

	\section{\label{summary}Summary and conclusions}
	
	Starting from the framework of scale invariant renormalisation, {we have shown that  equipping one Lagrangian with a dynamical renormalisation scale yields the same set of Green's functions as equipping an appropriately modified Lagrangian with a (traditional) constant renormalisation scale.} We have proposed three explicit examples of the construction of {this modified} Lagrangian at the one-loop level, and discussed how the construction can be ge\-ne\-ra\-li\-sed to any loop level.

	We have begun by considering a simple $\lambda h^4$ scalar theory accompanied by a second scalar field with a non-zero vacuum expectation value. Using this second scalar field {to dynamically generate} the re\-nor\-ma\-li\-sa\-tion scale, we preserved scale invariance of the theory at the quantum level. By requiring the one-loop level Green's functions {to come out equally}, we have shown that taking this dynamical renormalisation scale is equivalent to considering a new potential equipped with the vacuum expectation value of the dynamical scale as a renormalisation scale. This new potential contains an infinite number of new operators suppressed by powers of the vacuum expectation value of the dynamical scale. We have found that the dilatation symmetry of the new potential is hidden at the tree level, and {becomes} explicit again when quantum corrections are taken into account. We have done this construction explicitly at the one-loop level, and argued that it can be done at any loop level and for any potential.
	
	We then approached a scale invariant Yukawa model, containing a fermion and a scalar field with a non-zero vacuum expectation value. This model has a non-trivial wave-function renormalisation at the one-loop level, and permits a generalisation of the construction. Choo\-sing the scalar field as a dynamical renormalisation scale, we showed that {the one-loop quantum corrected theory remains scale invariant.} We then translated this renormalisation scale into a modification of the original Lagrangian, equipped with the scalar field vacuum expectation value as a renormalisation scale. We have shown that this new Lagrangian contains an infinite number of new operators suppressed by this vacuum expectation value, comparable to the first theory considered. Consequently, the new Lagrangian has its scale invariance hidden at the tree level, and recovers it at the one-loop level. We have argued that this construction can be done at any loop level, and for any theory admitting a perturbative expansion.
	
	We then considered an example in which this construction plays a role, namely a theory with gravity. We considered the standard model coupled in a non-minimal way to gravity. In this model, two prescriptions for re\-nor\-ma\-li\-sa\-tion appear, both of which naturally contain a dynamical renormalisation scale. We have translated these dynamical scales into choices of the tree potential of the theory, and found that choosing one or the other prescription amounts to adding an infinite series of operators suppressed by the Planck mass.
	
	{Taking the non-minimal coupling $\xi \gg 1$, this last example has a clear implication for Higgs inflation, in particular for the debate on frame independence. ``Frame independence" means that expressing one and the same theory in different frames yields identical predictions for physical observables. {Although we do not challenge frame independence}, we now clearly see that equipping a theory with two different renormalization scales (i.e., renormalizing a theory in one frame, or in the other at the same scale) yields two different QFTs, encoded in two different sets of Green's functions. Then, it does not come as a surprise that these two different theories yield different predictions for physical observables. }
	
	\section*{Acknowledgements}
	
	The authors thank  Mario Herrero-Valea, Alexander Monin, Kengo Shimada \RED{and an anonymous referee} for helpful discussions. This work was supported by the ERC-AdG-2015 grant 694896.  The work of M.S.  was supported partially by the Swiss National Science Foundation.

\end{document}